\documentclass[cameraready]{Interspeech}
\usepackage{multirow}
\title{Zero-VC: Zero-Lookahead Streaming Voice Conversion via Speaker Anonymization}

\author[affiliation={1,2}, orcid=0000-0002-6919-438X]{Yudong}{Li}
\author[affiliation={1}, orcid=0009-0003-0161-7176]{Zihao}{Fang}
\author[affiliation={1}, orcid=0009-0006-8000-9305]{Junwen}{Qiu}
\author[affiliation={3}, orcid=0000-0002-9528-7467]{Ruihai}{Jing}
\renewcommand{\authorsep}{, \\}
\author[affiliation={3}, orcid=0009-0000-6964-4830]{Ruixiang}{Hang}
\author[affiliation={1}, orcid=0009-0003-6718-7180]{Yingda}{Shen}
\author[affiliation={1,2,4}, orcid=0009-0001-1192-9857, correspondingauthor]{Zhizheng}{Wu}
\address{
    $^1$ The Chinese University of Hong Kong, Shenzhen \\
    $^2$ Shenzhen Loop Area Institute \\
    $^3$ Shenzhen Transsion Holdings Co., Ltd. \\
    $^4$ Amphion Technology Co.,Ltd.
}

\email{yudongli1@link.cuhk.edu.cn, wuzhizheng@cuhk.edu.cn}

\keywords{streaming voice conversion, speaker anonymization, zero-lookahead}

\usepackage{comment}

\begin{document}

\maketitle

% the abstract here must exactly match the abstract entered into the paper submission system
\begin{abstract}
% 1000 characters. ASCII characters only. No citations.
Streaming zero-shot voice conversion struggles to disentangle timbre from linguistic content without degrading utility or inflating latency. Current methods rely on information bottleneck (IB) or speaker perturbation. While IB filters out timbre, it discards prosody, forcing models to explicitly inject features like fundamental frequency. This often requires buffering future frames, creating algorithmic lookahead latency. On the other hand, existing perturbation methods largely overlook the crucial trade-off between timbre leakage and utility preservation. Recognizing this neglected trade-off, we find that the inherent objective of Speaker Anonymization (SA) aligns well with balancing these factors. Thus, we introduce SA as a novel perturbation mechanism to explicitly mitigate timbre leakage while retaining prosodic utility. Crucially, SA's robust representations significantly alleviate the generator's reliance on future context, enabling our strictly causal, zero-lookahead network. Audio samples are available at \url{https://amphionteam.github.io/Zero-VC-demo/}.
\end{abstract}

\section{Introduction}
Zero-shot Voice Conversion (VC) aims to convert a source speaker's voice to an unseen target speaker's 
voice using only a brief reference utterance \cite{sisman2020overview}. 
In recent years, driven by advances in deep learning, zero-shot VC has achieved remarkable success~\cite{zhangvevo,guo2025lscodec,ju2024naturalspeech,yang2024streamvc,liu2024zero,zhang2025vevo2,he2025noro,zhang2024leveraging}
and has been widely adopted in various real-world scenarios.
However, while non-streaming VC models have demonstrated impressive conversion quality, 
they require full-utterance input and are not feasible for real-time applications.
In real-time communication scenarios, streaming zero-shot VC 
faces a notoriously difficult challenge: effectively disentangling the source timbre from linguistic content 
while maintaining ultra-low processing latency. 

Current disentanglement strategies in streaming VC generally 
fall into two categories, each facing distinct limitations. 
The prevalent paradigm relies on Information Bottlenecks (IB)~\cite{yang2024streamvc,liu2025rt}. 
While extracting discrete speech units~\cite{yang2024streamvc} or articulatory features~\cite{liu2025rt} 
effectively filters out source identity, IB inherently acts as a destructive filter that 
discards fine-grained paralinguistic utility, notably prosody. 
To compensate, state-of-the-art streaming models explicitly extract and inject acoustic features such 
as fundamental frequency ($f_0$)~\cite{yang2024streamvc, liu2025rt}. 
However, calculating and smoothing these trajectories in a stable manner often necessitates temporal buffering. 
For instance, StreamVC requires a 
multi-frame context window to estimate $f_0$, which inherently introduces an algorithmic lookahead 
(e.g., 60~ms) and cascades into high end-to-end latency. This reliance on future context creates an 
architectural lower bound on latency, hindering hard-real-time deployments.

The second category is Speaker Perturbation~\cite{guo2025lscodec,liu2024zero}. 
Despite avoiding the destructive nature of strict bottlenecks, 
existing perturbation methods often struggle to achieve a favorable trade-off between timbre leakage and utility preservation. 
For example, LSCodec~\cite{guo2025lscodec} adopts signal processing techniques to perturb the source speaker's voice, 
but this method suffers from severe timbre leakage.
On the other hand, Seed-VC~\cite{liu2024zero} adopts an off-the-shelf VC model~\cite{qin2023openvoice} 
to perturb the source speaker's voice, which achieves a relatively balanced trade-off between timbre leakage and utility preservation.
However, because these methods are not explicitly designed to optimize this trade-off, the resulting balance remains sub-optimal.

In this paper, we recognize this neglected trade-off as the key to breaking the latency bottleneck. 
We find that the primary objective of Speaker Anonymization (SA)~\cite{panariello2024voiceprivacy}---concealing speaker identity while strictly 
preserving linguistic and prosodic utility---perfectly aligns with the ideal feature representation required by VC. 
Thus, we introduce SA into the streaming VC framework as an advanced speaker perturbation strategy. 
Unlike previous perturbation techniques, SA explicitly optimizes this trade-off, mitigating source timbre leakage 
while naturally preserving intact linguistic and prosodic dynamics.

% \footnote{Audio samples are available at  \url{https://amphionteam.github.io/Zero-VC-demo/}.}
Remarkably, the superior completeness of the SA-perturbed features yields a profound architectural advantage. 
Because the features retain natural prosody and temporal stability, the generative decoder's reliance on future frames 
is significantly alleviated. Exploiting this, we propose \textbf{Zero-VC}, a strictly causal, 
zero-lookahead (one-frame-in, one-frame-out) streaming architecture. 
While ablation studies indicate that adding a minor lookahead context yields marginal performance gains, 
our zero-lookahead configuration already achieves superior zero-shot VC quality. 
Leveraging SA, we successfully mitigate the fundamental trade-off between utility and leakage, 
thereby reducing latency to the single-frame theoretical minimum.

Our main contributions are summarized as follows:

\begin{figure*}[t]
  \centering
  \includegraphics[width=0.73\linewidth]{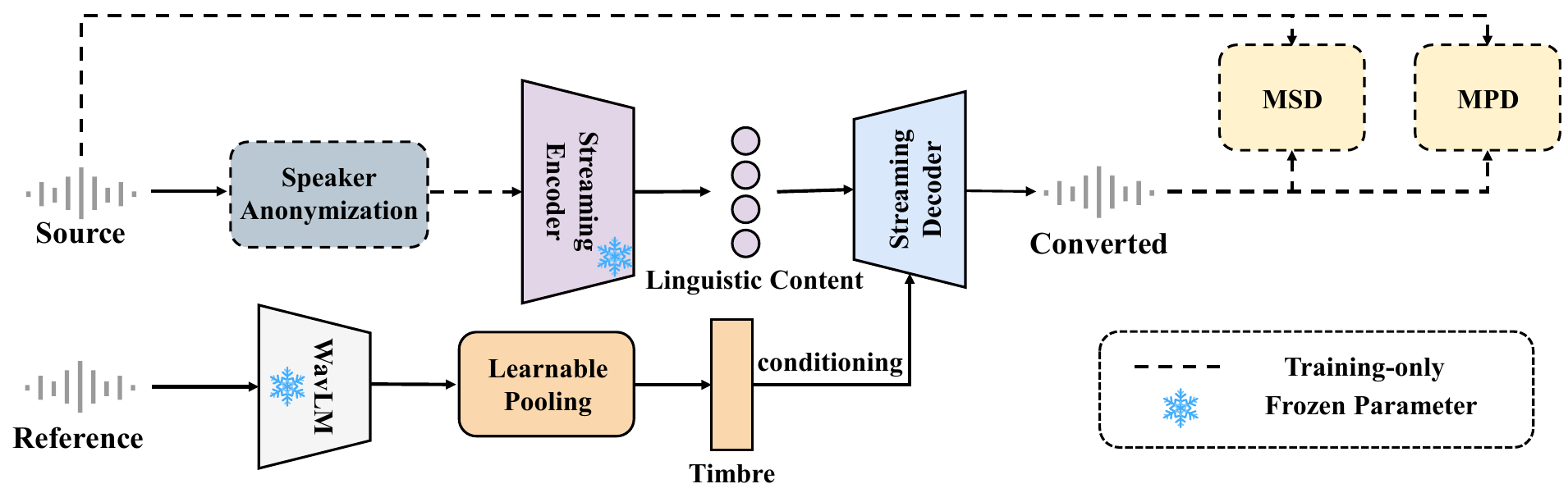}
  \caption{The overall framework of Zero-VC. During training, the same adversarial framework as HiFi-GAN~\cite{kong2020hifi} 
  is employed, utilizing a Multi-Scale Discriminator (MSD) and a Multi-Period Discriminator (MPD), 
  and source audio is processed by a Speaker Anonymization (SA) module to resolve the leakage-utility trade-off. 
  After training, the SA module, MSD, and MPD are discarded.}
  \label{fig:method}
\end{figure*}

\begin{itemize}
\item We identify that existing perturbation methods overlook the critical trade-off between timbre leakage and utility preservation. We introduce Speaker Anonymization (SA) as an effective perturbation mechanism, successfully mitigating timbre leakage while retaining essential prosodic utility.
\item We demonstrate that the highly informative SA representations significantly alleviate the model's reliance on future context. This enables the design of a strictly causal, zero-lookahead architecture, reducing algorithmic latency to a single frame (i.e., 20~ms), vastly outperforming the latency constraints of SOTA IB-based systems.
\item Comprehensive evaluations show that our zero-lookahead system achieves superior conversion quality and latency performance, establishing a robust balance between ultra-low latency, utility preservation, and timbre similarity.
\end{itemize}

\section{Method}
The overall framework of Zero-VC is illustrated in Fig.~\ref{fig:method}. 
Our model adopts a strictly causal, zero-lookahead architecture, consisting of a pretrained streaming encoder module (e.g., a distilled streaming w2v-bert-2.0~\cite{barrault2023seamless}), a timbre encoding module, and a streaming decoder.
The speaker anonymization (SA) module and two discriminators (i.e., MPD and MSD) shown in Fig.~\ref{fig:method} are used only for training and discarded during inference.

\subsection{SA-based Content Extraction}
To naturally disentangle linguistic content from the source speaker's identity, the source speech is first processed by an off-the-shelf SA module\footnote{\url{https://github.com/DigitalPhonetics/speaker-anonymization}}~\cite{meyer2024multilingual}. 
Unlike traditional lossy information bottlenecks, SA explicitly perturbs the speaker identity by mapping the source voice to a pseudo-speaker space while strictly preserving the temporal alignments, rich prosodic contours, and phonetic integrity.

The anonymized audio is then fed into a streaming encoder to extract linguistic content features. To satisfy the latency constraints, the encoder operates without any future lookahead, generating acoustic representations at a regular frame shift of 20~ms. By taking the SA-perturbed audio as input, the encoder yields linguistic features that are inherently devoid of the source speaker's timbre but retain highly expressive utility dynamics.

\subsection{Timbre Encoding}
To capture discriminative voice characteristics, 
we use the WavLM-large model\footnote{\url{https://github.com/microsoft/unilm/tree/master/wavlm}}~\cite{chen2022wavlm} 
to extract timbre features from reference speech, 
as it has proven to be highly effective for speaker verification tasks~\cite{yang2021superb}.
Specifically, we utilize the hidden states from the seventh transformer layer, denoted as $H = [h_1, h_2, \dots, h_L] \in \mathbb{R}^{L \times D}$, where $L$ is the sequence length and $D$ is the hidden dimension.
To aggregate these frame-level representations into a single, fixed-length speaker embedding ($T=1$), we apply an attention-based learnable pooling layer. The attention weight $\alpha_i$ for each frame is computed via a learnable linear projection $W_p$, and the global timbre condition $c_{timbre} \in \mathbb{R}^{D}$ is then obtained by a weighted sum of the temporal features:
\begin{align}
    \alpha_i = \frac{\exp(W_p h_i)}{\sum_{j=1}^{L} \exp(W_p h_j)},\quad c_{timbre} = \sum_{i=1}^{L} \alpha_i h_i.
\end{align}

This mechanism allows the network to automatically attend to the most speaker-discriminative phonetic segments within the reference speech.

% Then, we use the outputs of the seventh layer and then pass them through a learnable pooling layer, 
% which pools features along the time dimension to aggregate them into a single, fixed-length speaker embedding ($T=1$). 
% This embedding serves as the global timbre condition for the downstream decoder.

\subsection{Streaming Decoder and Training Objectives}
The decoder architecture is mainly based on HiFi-GAN~\cite{kong2020hifi}, 
but with critical modifications for zero-lookahead streaming. 
All standard convolutions are replaced with causal convolutions to 
ensure that the generation of the current frame does not depend on any future context.
% All standard convolutions within the residual blocks are replaced with strictly causal convolutions. Mathematically, given an intermediate feature sequence $x_{1:T}$, the output at time step $t$ for a causal convolution layer with kernel size $K$ and dilation rate $d$ is computed strictly using elements $\{x_{t - d \cdot i}\}_{i=0}^{K-1}$. This is implemented by applying an asymmetric zero-padding of size $d \times (K-1)$ exclusively to the left (past) side of the temporal dimension. This ensures that the generation of the current frame does not depend on any future context, satisfying the hard constraints of real-time streaming.

Following~\cite{kashkin2022hifi}, we employ a three-layer Conv1D to inject the global timbre embedding $c_{timbre}$ into the generation process. Given an intermediate feature map $x$, the conditioning layer shifts $x$ by an offset predicted by the convolutional layers:
\begin{align}
    x' =  x + \text{Convs}(c_{timbre}),
\end{align}
where $\text{Convs}()$ denotes a stack of convolutional layers.

During training, we employ the same adversarial framework as HiFi-GAN~\cite{kong2020hifi}, utilizing a 
Multi-Scale Discriminator (MSD) and a Multi-Period Discriminator (MPD). 
The overall training objective is a weighted sum of the Mel-Spectrogram loss ($\mathcal{L}_{mel}$), 
the feature matching loss ($\mathcal{L}_{fm}$), and the GAN loss ($\mathcal{L}_{adv}$). 
Specifically, the final objectives for the generator and discriminator are formulated as:

\begin{align}
  \mathcal{L}_{G} = \lambda_{mel}\mathcal{L}_{mel} &+ \sum^K_{k=1}[\lambda_{fm}\mathcal{L}_{fm}(G;D_k) + \lambda_{adv}\mathcal{L}_{adv}(G;D_k)], \nonumber \\
  \mathcal{L}_{D} &= \sum^K_{k=1}\mathcal{L}_{adv}(D_k;G),
  \label{eq:Gloss}
\end{align}

% \begin{align}
  
%   \label{eq:Dloss}
% \end{align}
where $D_k$ denotes the $k$-th sub-discriminator in the MPD and MSD, and $G$ denotes the generator.

\subsection{Streaming Inference Strategy}
A key advantage of Zero-VC lies in its highly efficient streaming inference. Because the architecture is strictly causal, the model processes the incoming audio chunk-by-chunk (i.e., every 20~ms). During inference, we maintain a state buffer (cache) for the causal convolutional layers, which stores only the required past frames defined by the receptive field. As a result, the computational complexity per frame remains constant $O(1)$, independent of the total sequence length. This streaming cache mechanism, combined with the zero-lookahead design, enables Zero-VC to achieve consistently minimal algorithmic latency.

\section{Experiments}
\subsection{Experimental Setup}
\subsubsection{Datasets}
Our model is trained on LibriTTS~\cite{zen2019libritts}, which is an English corpus with 585 hours of speech data. 
We discard utterances shorter than 4 seconds and resample all audio to 16~kHz, 
resulting in approximately 460 hours of training data. 
For evaluation, we utilize the English subset of the seed-tts-eval dataset~\cite{anastassiou2024seed},
comprising approximately 1,000 pairs of samples from the Common Voice dataset~\cite{ardila2020common}.

\begin{figure}[t]
  \centering
  \includegraphics[width=0.8\linewidth]{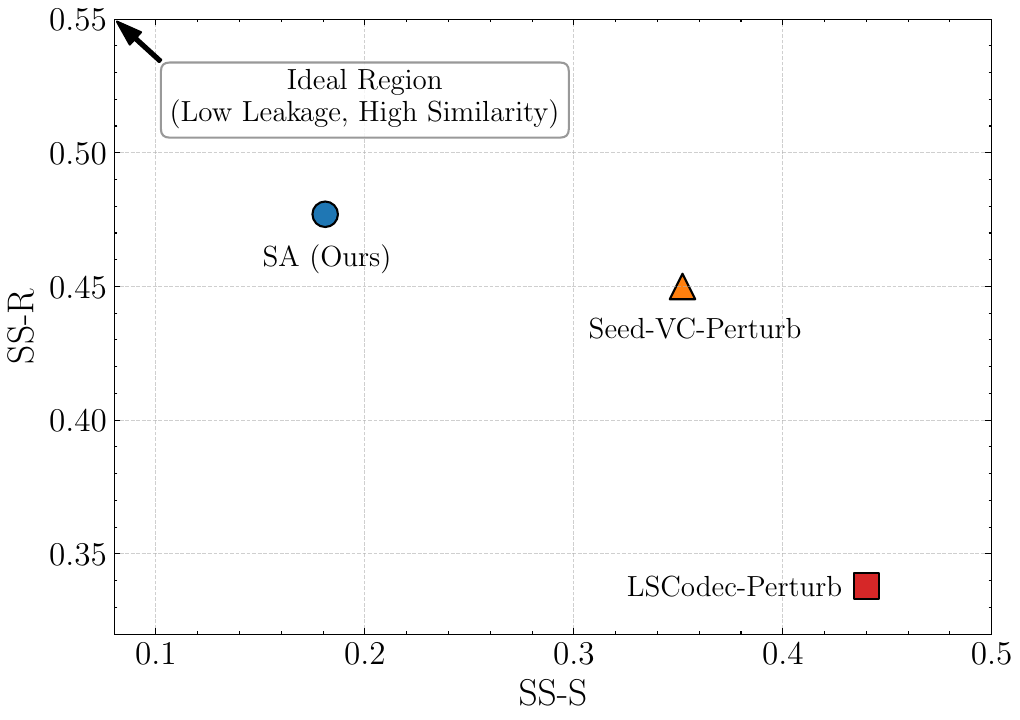}
  \caption{Speaker similarity trade-off. SS-S and SS-R denote speaker similarity to source and reference, respectively.
  Proximity to the top-left region signifies a more favorable trade-off.}
  \label{fig:tradeoff}
\end{figure}

\subsubsection{Evaluation metrics}
We evaluate Speaker Similarity (SS) using a WavLM-large model fine-tuned on the speaker verification task~\cite{chen2022large} to extract speaker embeddings. These embeddings are used to calculate the cosine similarity between the converted speech and the corresponding source or reference speech. We report both the similarity to the source speech (SS-S, where a lower value indicates less leakage) and the reference speech (SS-R, where higher is better). 
Word Error Rate (WER) is computed using Whisper-large-v3~\cite{radford2023robust} to 
measure intelligibility. Prosody preservation is measured by F0 Pearson Coefficients (FPC). 
An overall quality score (OVRL) is predicted using Microsoft's DNSMOS P.835 model~\cite{reddy2022dnsmos} 
(higher scores indicate better perceived speech quality). For subjective evaluation, we measure naturalness mean 
opinion score (NMOS) and speaker similarity mean opinion score (SMOS). 
NMOS considers intelligibility, prosody, and audio quality, while SMOS evaluates speaker similarity from the 
perspective of human perception.
All subjective metrics are reported with their 95\% confidence intervals.

\subsubsection{Implementation Details}
The VC model is trained for 1.2M steps for final evaluation, 
while ablation models are trained for 120k steps. 
We use the AdamW optimizer~\cite{loshchilovdecoupled} with $\beta_1 = 0.8$, $\beta_2 = 0.99$, 
a learning rate of $6 \times 10^{-4}$ and a weight decay of 0.01, 
coupled with a Cosine-Annealing scheduler~\cite{loshchilov2017sgdr}. The batch size is set to 30, 
where each batch consists of a 2-second source segment and a 2-second reference segment. 
The loss weights are configured as $\lambda_{fm} = 3$, $\lambda_{mel} = 51$, 
and $\lambda_{adv} = 1$.

\subsection{Ablation Studies}
\begin{table}[th]
    \caption{Objective evaluation of intermediate audio generated by different perturbation strategies. \textbf{Boldface} denotes the best result, while \underline{underlining} indicates the second best.}
    \label{tab:processing}
    \centering
    \resizebox{1.0\columnwidth}{!}{
        \begin{tabular}{c|cccc}
          \hline
          Method & SS-S & WER(\%) & FPC & OVRL \\
          \hline
          LSCodec-Perturb~\cite{guo2025lscodec}                       & 0.704 & \textbf{2.15} & \textbf{0.891} & 3.054             \\
          Seed-VC-Perturb~\cite{liu2024zero}                       & \underline{0.411} & \underline{4.45} & 0.688 & \textbf{3.249}              \\
          SA                            & \textbf{0.119} & 8.33 & \underline{0.718} & \underline{3.175}      \\
          \hline
        \end{tabular}
    }
  
\end{table}

To validate the superiority of SA over existing perturbation methods, 
we conduct ablation studies from three perspectives. 

First, we directly evaluate the perturbed audio (Tab.~\ref{tab:processing}). 
Note that for baseline perturbation methods, 
we use the specific perturbation modules utilized in LSCodec~\cite{guo2025lscodec} and Seed-VC~\cite{liu2024zero} to process audio, 
rather than evaluating their entire voice conversion pipelines.
LSCodec-Perturb exhibits severe timbre leakage with an SS-S of 0.704. 
While Seed-VC-Perturb reduces this leakage (SS-S 0.411), 
it causes a drop in FPC (0.688). 
Our SA method effectively neutralizes the source timbre (achieving an exceptionally low SS-S of 0.119) 
and better preserves the original prosodic contour (FPC 0.718). 
Although the intermediate WER of SA appears relatively high at this stage, 
downstream training (Tab.~\ref{tab:training}) reveals that the model can still achieve considerable intelligibility, 
demonstrating that SA provides a highly robust foundation for disentanglement.

% \begin{figure}[t]
%   \centering
%   \includegraphics[width=0.8\linewidth]{utility_scatter.pdf}
%   \caption{Utility preservation trade-off.}
%   \label{fig:utility}
% \end{figure}

\begin{table}[th]
    \caption{Utility preservation trade-off of models trained with different perturbation strategies. 
    \textbf{Boldface} denotes the best result, while \underline{underlining} indicates the second best.}
    \label{tab:training}
    \centering
    \begin{tabular}{c|ccc}
      \hline
      Method & WER & FPC & OVRL \\
      \hline
      LSCodec-Perturb                       & \textbf{2.64} & \textbf{0.681} & \textbf{3.097}             \\
      Seed-VC-Perturb                       & \underline{2.67} & 0.659 & \underline{3.093}              \\
      SA                            & 3.82 & \underline{0.671} & 3.040      \\
      \hline
    \end{tabular}
  
\end{table}

\begin{figure}[t]
  \centering
  \includegraphics[width=0.8\linewidth]{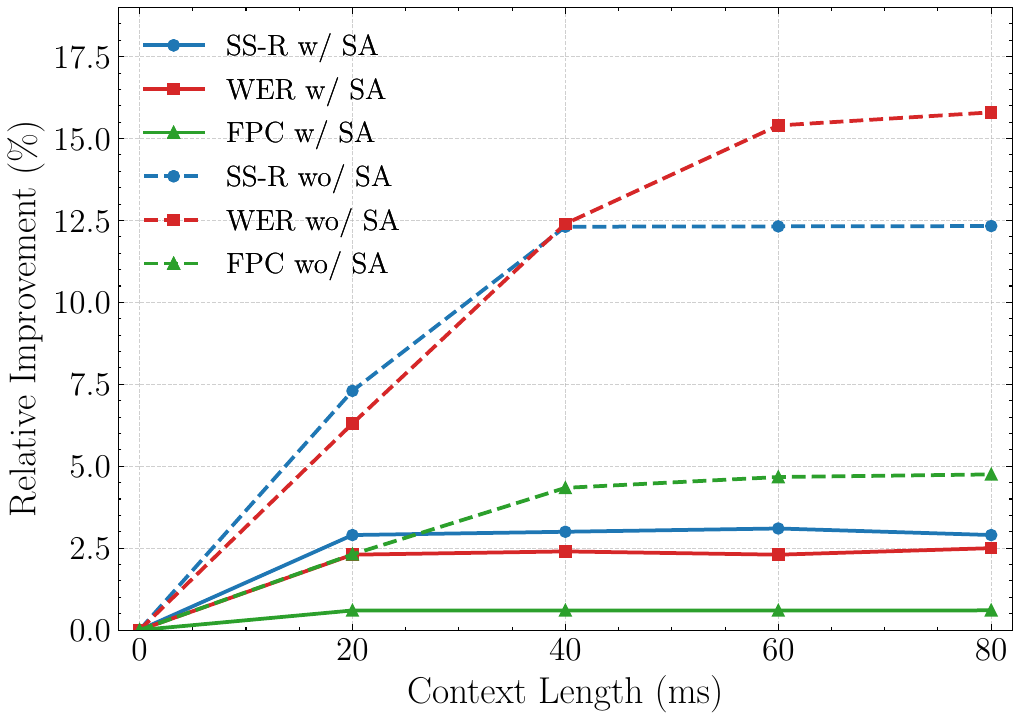}
  \caption{Relative improvement brought by lookahead context. 
  Solid lines denote the models with SA perturbation, and dashed lines denote the models without SA perturbation.}
  \label{fig:improvement}
\end{figure}

\begin{table*}[th]
    \caption{Zero-Shot VC performance. RTF is tested on an Intel Xeon Platinum 8468V-2.4~GHz CPU. \textbf{Boldface} denotes the best result, while \underline{underlining} indicates the second best. NMOS and SMOS are reported with their 95\% confidence intervals.}
    \label{tab:zero-shot}
    \centering
    \begin{tabular}{c|c|ccccccccc}
      \hline
      Group &Method & SS-S & SS-R & SMOS & WER(\%) & FPC & OVRL & NMOS & RTF \\
      \hline
      % \multirow{4}{*}{Non-Streaming} & FACodec~\cite{ju2024naturalspeech} & 0.341 & 0.365 & 3.740 & 0.6664 & 3.0599 & 0.0672 \\
      \multirow{3}{*}{Non-Streaming} & LSCodec~\cite{guo2025lscodec} & \underline{0.277} & 0.426 & 3.64$\pm$0.07 & 9.00 & 0.650 & 3.116 & 3.70$\pm$0.06 & \underline{0.077} \\
      % & LSCodec~\cite{guo2025lscodec} & 0.277 & 0.426 & 9.000 & 0.6503 & 3.1163 & 0.0767 \\
      & CosyVoice~\cite{du2024cosyvoice} & 0.313 & \underline{0.502} & \underline{3.78$\pm$0.06} & 4.02 & 0.644 & \textbf{3.182} & \textbf{3.82}$\pm$\textbf{0.05}  & 2.441 \\
      & Seed-VC-Small~\cite{liu2024zero} & 0.402 & 0.415 & 3.62$\pm$0.09 & \textbf{2.47} & \underline{0.661} & \underline{3.141} & 3.77$\pm$0.06 & 0.508 \\

      \hline
      % \multirow{2}{*}{} & CosyVoice2 & 0.294 & \textbf{0.522} & 4.202 & 0.6306 & 3.1516 & 0.6014 \\
      Streaming & Zero-VC & \textbf{0.171} & \textbf{0.521} & \textbf{3.88}$\pm$\textbf{0.05} & \underline{3.96} & \textbf{0.688} & 3.044 & \underline{3.81$\pm$0.07} & \textbf{0.063}      \\
      \hline
    \end{tabular}
  
\end{table*}

\begin{table}[th]
  \caption{Algorithmic latency comparison. Note that algorithmic latency does not include inference latency.}
  \label{tab:latency}
  \centering
  \resizebox{1.0\columnwidth}{!}{
    \begin{tabular}{c|ccccc}
      \hline
      Method & DualVC3~\cite{ning2024dualvc} & StreamVC~\cite{yang2024streamvc} & RT-VC~\cite{liu2025rt} & Zero-VC \\
      \hline
      Latency(ms)                       & 40 & 60 & 47 & \textbf{20}             \\
      \hline
    \end{tabular}
  }

\end{table}

% \begin{table}[th]
%     \caption{Comparison of different perturbation methods used for training VC models}
%     \label{tab:training}
%     \centering
%     \begin{tabular}{c|ccccc}
%       \hline
%       Method & SS-S & SS-R & WER & FPC & OVRL \\
%       \hline
%       LSCodec                       & 0.431 & 0.401 & \textbf{2.638} & \textbf{0.6808} & \textbf{3.0969}             \\
%       Seed-VC                       & \underline{0.352} & \underline{0.450} & \underline{2.670} & 0.6587 & \underline{3.0933}              \\
%       SA                            & \textbf{0.181} & \textbf{0.477} & 3.821 & \underline{0.6709} & 3.0395      \\
%       \hline
%     \end{tabular}
  
% \end{table}
Second, we train identical VC models using the three perturbation strategies to analyze the multi-dimensional trade-offs. As shown in Fig.~\ref{fig:tradeoff}, the model equipped with SA (blue circle) resides closest to the ideal region for speaker similarity, achieving the highest target similarity (SS-R) while exhibiting the lowest source timbre leakage (SS-S). In terms of utility preservation (Tab.~\ref{tab:training}), it is crucial to note that while LSCodec-Perturb strictly preserves prosody (achieving a high FPC), this metric becomes less meaningful when SS-R is extremely low. In such cases, the model may simply reconstruct the source speech without effectively altering the timbre. Furthermore, as indicated by the OVRL scores, the SA-based model maintains a competitive overall speech naturalness (OVRL 3.040) that is closely comparable to the baseline methods. Ultimately, our SA-based model strikes a highly competitive balance, maintaining strong intelligibility, prosody, and overall quality without sacrificing the core objective of high-fidelity timbre conversion.

Finally, to prove our hypothesis that SA alleviates the need for future context, 
we evaluate the relative improvement of models given varying lookahead lengths (Fig.~\ref{fig:improvement}). 
Strikingly, the performance metrics (SS-R, WER, and FPC) of the SA-trained model (solid lines) 
saturate almost immediately at 0 to 20~ms, showing less than a 3\% relative improvement 
even when provided with up to 80~ms of future context. Conversely, 
the model without SA (dashed lines) exhibits a strong dependency on lookahead, 
requiring at least 40 to 60~ms of future frames to stabilize its WER and SS-R improvements 
(gaining more than 12\% to 15\%). This stark contrast compellingly verifies that the high-quality, 
prosody-rich representations provided by SA inherently alleviate 
the generator's reliance on future context, making a zero-lookahead architecture highly effective.

\subsection{Evaluation on Zero-Shot VC}
Conducting a fair comparison against state-of-the-art (SOTA) real-time streaming VC models presents a challenge, 
as the majority of highly optimized models (e.g., StreamVC~\cite{yang2024streamvc}, RT-VC~\cite{liu2025rt}) remain closed-source. 
Therefore, to comprehensively assess Zero-VC, we compare its conversion quality against prominent 
open-source non-streaming VC systems, 
and evaluate its algorithmic latency against reported metrics from recent streaming SOTA models.

As shown in Tab.~\ref{tab:zero-shot}, despite operating in a strict 
zero-lookahead streaming paradigm, Zero-VC demonstrates highly competitive, 
and in many aspects superior, conversion capabilities compared to heavy non-streaming models. 
Objective metrics reveal that Zero-VC achieves the lowest source leakage (SS-S 0.171) and 
the highest target similarity (SS-R 0.521). This superiority is strongly corroborated by 
our subjective evaluations: Zero-VC attains the highest SMOS ($3.88 \pm 0.05$), proving its 
exceptional zero-shot speaker adaptation capability.
Regarding utility and speech quality, Zero-VC demonstrates remarkable robustness. Impressively, it obtains the highest FPC of 0.688 and ranks second best in both intelligibility (achieving a WER of 3.96\%, trailing only Seed-VC-Small) and perceptual naturalness (attaining an NMOS of $3.81 \pm 0.07$, which is only marginally behind CosyVoice). 
% This clearly indicates that the strict zero-lookahead constraint does not degrade perceptual naturalness or linguistic integrity, largely thanks to the rich utility retained by the SA features.
Notably, Zero-VC is lightweight, 
achieving a Real-Time Factor (RTF) of 0.063 on an Intel Xeon Platinum 8468V-2.4~GHz CPU, 
ensuring it comfortably meets real-time processing constraints.

Another primary advantage of our SA-driven architecture is the alleviation of dependency on future context.
Tab.~\ref{tab:latency} illustrates the algorithmic latency of Zero-VC compared to recent streaming solutions. 
Note that the algorithmic latency does not include inference latency for a fair comparison, since
other models are closed-source.
By strictly operating on a 20~ms frame basis without requiring future frame buffers for 
acoustic feature smoothing, 
Zero-VC achieves a minimal algorithmic latency of 20~ms. 
This significantly undercuts the 40-60~ms latency barriers typical of IB-based models, 
setting a new benchmark for hard-real-time streaming voice conversion.

\section{Limitations and Future Work}
While Zero-VC achieves superior zero-lookahead latency for the core VC generator, the current training pipeline relies on an off-the-shelf SA module to preprocess source audio during training. Depending on the SA module's specific implementation, this pre-processing step may introduce its own training overhead. Future work will focus on integrating the SA objective directly into the training paradigm in an end-to-end manner, further optimizing total system latency, and extending the framework to support cross-lingual zero-shot voice conversion.

\section{Conclusion}
In this paper, we present Zero-VC, a novel strictly causal, zero-lookahead streaming voice conversion system. 
By identifying the critical trade-off between timbre leakage and utility preservation, 
we introduce Speaker Anonymization as an innovative perturbation mechanism. 
This effective feature disentanglement significantly alleviates the decoder's reliance on future context. 
Experimental results demonstrate that our SA-based approach outperforms traditional perturbation methods 
in minimizing source leakage and maximizing target similarity, ultimately paving the way for ultra-low latency, 
high-fidelity real-time voice conversion.

\section{Acknowledgments}
This work is partially supported by the Internal Project Fund from Shenzhen Research Institute of Big Data (Grant No. T00120230002) and the Program for Guangdong Introducing Innovative and Enterpreneurial Teams (Grant No. 2023ZT10X044). We thank the anonymous reviewers for their insightful comments and suggestions. We appreciate the efforts of all the subjects during the subjective evaluation.

\section{Generative AI Use Disclosure}
The authors used Google Gemini for language polishing and grammar checking. After polishing, the manuscript was revised, and the authors take full responsibility for the originality and final content of the publication.

\bibliographystyle{IEEEtran}
\bibliography{mybib}

@article{sisman2020overview,
  title={An overview of voice conversion and its challenges: From statistical modeling to deep learning},
  author={Sisman, Berrak and Yamagishi, Junichi and King, Simon and Li, Haizhou},
  journal={IEEE/ACM transactions on audio, speech, and language processing},
  volume={29},
  pages={132--157},
  year={2020},
  publisher={IEEE}
}

@inproceedings{yang2024streamvc,
  title={Streamvc: Real-time low-latency voice conversion},
  author={Yang, Yang and Kartynnik, Yury and Li, Yunpeng and Tang, Jiuqiang and Li, Xing and Sung, George and Grundmann, Matthias},
  booktitle={ICASSP 2024-2024 IEEE International Conference on Acoustics, Speech and Signal Processing (ICASSP)},
  pages={11016--11020},
  year={2024},
  organization={IEEE}
}

@inproceedings{liu2025rt,
  title={RT-VC: Real-Time Zero-Shot Voice Conversion with Speech Articulatory Coding},
  author={Liu, Yisi and Wang, Chenyang and Kim, Hanjo and Khan, Raniya and Anumanchipalli, Gopala},
  booktitle={Proceedings of the 63rd Annual Meeting of the Association for Computational Linguistics (Volume 3: System Demonstrations)},
  pages={385--393},
  year={2025}
}

@inproceedings{guo2025lscodec,
  title={LSCodec: Low-Bitrate and Speaker-Decoupled Discrete Speech Codec},
  author={Guo, Yiwei and Li, Zhihan and Du, Chenpeng and Wang, Hankun and Chen, Xie and Yu, Kai},
  booktitle={Proc. Interspeech 2025},
  pages={5018--5022},
  year={2025}
}

@article{liu2024zero,
  title={Zero-shot voice conversion with diffusion transformers},
  author={Liu, Songting},
  journal={arXiv preprint arXiv:2411.09943},
  year={2024}
}

@article{qin2023openvoice,
  title={Openvoice: Versatile instant voice cloning},
  author={Qin, Zengyi and Zhao, Wenliang and Yu, Xumin and Sun, Xin},
  journal={arXiv preprint arXiv:2312.01479},
  year={2023}
}

@inproceedings{zhangvevo,
  title={Vevo: Controllable Zero-Shot Voice Imitation with Self-Supervised Disentanglement},
  author={Zhang, Xueyao and Zhang, Xiaohui and Peng, Kainan and Tang, Zhenyu and Manohar, Vimal and Liu, Yingru and Hwang, Jeff and Li, Dangna and Wang, Yuhao and Chan, Julian and others},
  booktitle={The Thirteenth International Conference on Learning Representations}
}

@inproceedings{ju2024naturalspeech,
  title={NaturalSpeech 3: zero-shot speech synthesis with factorized codec and diffusion models},
  author={Ju, Zeqian and Wang, Yuancheng and Shen, Kai and Tan, Xu and Xin, Detai and Yang, Dongchao and Liu, Yanqing and Leng, Yichong and Song, Kaitao and Tang, Siliang and others},
  booktitle={Proceedings of the 41st International Conference on Machine Learning},
  pages={22605--22623},
  year={2024}
}

@inproceedings{ning2024dualvc,
  title={DualVC 3: Leveraging Language Model Generated Pseudo Context for End-to-end Low Latency Streaming Voice Conversion},
  author={Ning, Ziqian and Wang, Shuai and Zhu, Pengcheng and Wang, Zhichao and Yao, Jixun and Xie, Lei and Bi, Mengxiao},
  booktitle={Proc. Interspeech 2024},
  pages={197--201},
  year={2024}
}

@article{panariello2024voiceprivacy,
  title={The VoicePrivacy 2022 Challenge: Progress and Perspectives in Voice Anonymisation},
  author={Panariello, Michele and Tomashenko, Natalia and Wang, Xin and Miao, Xiaoxiao and Champion, Pierre and Nourtel, Hubert and Todisco, Massimiliano and Evans, Nicholas and Vincent, Emmanuel and Yamagishi, Junichi},
  journal={IEEE/ACM Transactions on Audio, Speech, and Language Processing},
  volume={32},
  pages={3477--3491},
  year={2024},
  publisher={Institute of Electrical and Electronics Engineers (IEEE)}
}

@inproceedings{meyer2024multilingual,
  title     = {Probing the Feasibility of Multilingual Speaker Anonymization},
  author    = {Sarina Meyer and Florian Lux and Ngoc Thang Vu},
  year      = {2024},
  booktitle = {Interspeech 2024},
  pages     = {4448--4452},
  doi       = {10.21437/Interspeech.2024-1615},
  issn      = {2958-1796},
}

@article{chen2022wavlm,
  title={Wavlm: Large-scale self-supervised pre-training for full stack speech processing},
  author={Chen, Sanyuan and Wang, Chengyi and Chen, Zhengyang and Wu, Yu and Liu, Shujie and Chen, Zhuo and Li, Jinyu and Kanda, Naoyuki and Yoshioka, Takuya and Xiao, Xiong and others},
  journal={IEEE Journal of Selected Topics in Signal Processing},
  volume={16},
  number={6},
  pages={1505--1518},
  year={2022},
  publisher={IEEE}
}

@inproceedings{yang2021superb,
  title={SUPERB: Speech Processing Universal PERformance Benchmark},
  author={Yang, Shu-wen and Chi, Po-Han and Chuang, Yung-Sung and Lai, Cheng-I Jeff and Lakhotia, Kushal and Lin, Yist Y and Liu, Andy T and Shi, Jiatong and Chang, Xuankai and Lin, Guan-Ting and others},
  booktitle={Proc. Interspeech 2021},
  pages={1194--1198},
  year={2021}
}

@article{kashkin2022hifi,
  title={Hifi-vc: High quality asr-based voice conversion},
  journal={arXiv preprint arXiv:2203.16937},
  year={2022}
}

@article{kong2020hifi,
  title={Hifi-gan: Generative adversarial networks for efficient and high fidelity speech synthesis},
  author={Kong, Jungil and Kim, Jaehyeon and Bae, Jaekyoung},
  journal={Advances in neural information processing systems},
  volume={33},
  pages={17022--17033},
  year={2020}
}

@inproceedings{zen2019libritts,
  title={LibriTTS: A Corpus Derived from LibriSpeech for Text-to-Speech},
  author={Zen, Heiga and Dang, Viet and Clark, Rob and Zhang, Yu and Weiss, Ron J and Jia, Ye and Chen, Zhifeng and Wu, Yonghui},
  booktitle={Proc. Interspeech 2019},
  pages={1526--1530},
  year={2019}
}

@article{anastassiou2024seed,
  title={Seed-tts: A family of high-quality versatile speech generation models},
  author={Anastassiou, Philip and Chen, Jiawei and Chen, Jitong and Chen, Yuanzhe and Chen, Zhuo and Chen, Ziyi and Cong, Jian and Deng, Lelai and Ding, Chuang and Gao, Lu and others},
  journal={arXiv preprint arXiv:2406.02430},
  year={2024}
}

@inproceedings{radford2023robust,
  title={Robust speech recognition via large-scale weak supervision},
  author={Radford, Alec and Kim, Jong Wook and Xu, Tao and Brockman, Greg and McLeavey, Christine and Sutskever, Ilya},
  booktitle={International conference on machine learning},
  pages={28492--28518},
  year={2023},
  organization={PMLR}
}

@inproceedings{chen2022large,
  title={Large-scale self-supervised speech representation learning for automatic speaker verification},
  author={Chen, Zhengyang and Chen, Sanyuan and Wu, Yu and Qian, Yao and Wang, Chengyi and Liu, Shujie and Qian, Yanmin and Zeng, Michael},
  booktitle={ICASSP 2022-2022 IEEE International Conference on Acoustics, Speech and Signal Processing (ICASSP)},
  pages={6147--6151},
  year={2022},
  organization={IEEE}
}

@inproceedings{reddy2022dnsmos,
  title={DNSMOS P. 835: A non-intrusive perceptual objective speech quality metric to evaluate noise suppressors},
  author={Reddy, Chandan KA and Gopal, Vishak and Cutler, Ross},
  booktitle={ICASSP 2022-2022 IEEE international conference on acoustics, speech and signal processing (ICASSP)},
  pages={886--890},
  year={2022},
  organization={IEEE}
}

@inproceedings{loshchilovdecoupled,
  title={Decoupled Weight Decay Regularization},
  author={Loshchilov, Ilya and Hutter, Frank},
  booktitle={International Conference on Learning Representations}
}

@inproceedings{loshchilov2017sgdr,
  title={SGDR: Stochastic Gradient Descent with Warm Restarts},
  author={Loshchilov, Ilya and Hutter, Frank},
  booktitle={International Conference on Learning Representations},
  year={2017}
}

@article{du2024cosyvoice,
  title={Cosyvoice: A scalable multilingual zero-shot text-to-speech synthesizer based on supervised semantic tokens},
  author={Du, Zhihao and Chen, Qian and Zhang, Shiliang and Hu, Kai and Lu, Heng and Yang, Yexin and Hu, Hangrui and Zheng, Siqi and Gu, Yue and Ma, Ziyang and others},
  journal={arXiv preprint arXiv:2407.05407},
  year={2024}
}

@inproceedings{ardila2020common,
  title={Common voice: A massively-multilingual speech corpus},
  author={Ardila, Rosana and Branson, Megan and Davis, Kelly and Kohler, Michael and Meyer, Josh and Henretty, Michael and Morais, Reuben and Saunders, Lindsay and Tyers, Francis and Weber, Gregor},
  booktitle={Proceedings of the twelfth language resources and evaluation conference},
  pages={4218--4222},
  year={2020}
}

@article{zhang2025vevo2,
  title={Vevo2: Bridging controllable speech and singing voice generation via unified prosody learning},
  author={Zhang, Xueyao and Zhang, Junan and Wang, Yuancheng and Wang, Chaoren and Chen, Yuanzhe and Jia, Dongya and Chen, Zhuo and Wu, Zhizheng},
  journal={arXiv e-prints},
  pages={arXiv--2508},
  year={2025}
}

@inproceedings{he2025noro,
  title={Noro: Noise-Robust One-Shot Voice Conversion with Hidden Speaker Representation Learning},
  author={He, Haorui and Song, Yuchen and Wang, Yuancheng and Li, Haoyang and Zhang, Xueyao and Wang, Li and Huang, Gongping and Chng, Eng Siong and Wu, Zhizheng},
  booktitle={2025 Asia Pacific Signal and Information Processing Association Annual Summit and Conference (APSIPA ASC)},
  pages={2247--2251},
  year={2025},
  organization={IEEE}
}

@inproceedings{zhang2024leveraging,
  title={Leveraging diverse semantic-based audio pretrained models for singing voice conversion},
  author={Zhang, Xueyao and Fang, Zihao and Gu, Yicheng and Chen, Haopeng and Zou, Lexiao and Zhang, Junan and Xue, Liumeng and Wu, Zhizheng},
  booktitle={2024 IEEE Spoken Language Technology Workshop (SLT)},
  pages={758--765},
  year={2024},
  organization={IEEE}
}

@article{barrault2023seamless,
  title={Seamless: Multilingual Expressive and Streaming Speech Translation},
  author={Barrault, Lo{\"\i}c and Chung, Yu-An and Meglioli, Mariano Coria and Dale, David and Dong, Ning and Duppenthaler, Mark and Duquenne, Paul-Ambroise and Ellis, Brian and Elsahar, Hady and Haaheim, Justin and others},
  journal={arXiv preprint arXiv:2312.05187},
  year={2023}
}

\end{document}